\documentclass[reprint, showpacs, superscriptaddress, aps, pre, twocolumn, 10pt]{revtex4-1}

\usepackage{graphicx} 
\usepackage{dcolumn}
\usepackage{bm}
\usepackage{amssymb}
\usepackage{amsmath}
\usepackage{wasysym}
\usepackage{color}
\usepackage{hyperref}
\usepackage{dcolumn}
\usepackage{float}
\usepackage{flushend}
\usepackage{multirow}
\usepackage{cancel}
\hypersetup{
colorlinks,
citecolor=blue,
filecolor=blue,
linkcolor=blue,
urlcolor=blue}

\newcommand{\be}{\begin{equation}}
\newcommand{\ee}{\end{equation}}
\newcommand{\bse}{\begin{subequations}}
\newcommand{\ese}{\end{subequations}}
\newcommand{\bea}{\begin{eqnarray}}
\newcommand{\eea}{\end{eqnarray}}
\newcommand{\ba}{\begin{array}}
\newcommand{\ea}{\end{array}}

\begin{document}
\title{Turbulent drag reduction: A universal perspective from energy fluxes}
\author{Mahendra K. Verma}
\email{mkv@iitk.ac.in}
\affiliation {Department of Physics, Indian Institute of Technology, Kanpur, India 208016}


\begin{abstract} 
Injection of dilute polymer in a turbulent flow suppresses frictional drag.  This challenging  and technologically important problem remains primarily unresolved due to the complex nature of the flow.  An important factor in the drag reduction is the energy transfer from the velocity field to the polymers.  In this paper we quantify this process using energy fluxes, as well as show its universality in diverse flows such as magnetohydrodynamics,  quasi-static magnetohydrodynamics, and bubbly turbulence. We show that in such flows, the  transfer from kinetic energy to elastic energy leads to a reduction in kinetic energy flux compared to the corresponding hydrodynamic turbulence. This leads to a reduction in nonlinearity of the velocity field that  results into a more ordered flow  and a suppression of turbulent drag.

\end{abstract}

\maketitle
 

Frictional force in  a turbulent flow   is proportional to  square of the flow velocity~\cite{Davidson:book:Turbulence,Sagaut:book}.  This  steep  dependence of frictional force or {\em turbulent drag} on velocity makes it a major  challenge in aerospace and automobile industry, as well as for flow engineering.  Hence, it is an important area of research.  In this paper we  address this problem in a general framework of energy transfers and fluxes.  In addition to application of energy transfers  to drag reduction in polymeric turbulence, we make an unexpected prediction that magnetohydrodynamics and quasi-static magnetohydrodynamics too exhibit turbulent drag reduction.   We show that an inclusion of magnetic field or polymers in a turbulent flow leads to reductions of kinetic energy cascade rate,  nonlinearity, and turbulent drag.

Past experiments and numerical simulation  reported turbulent drag reduction in solution with dilute polymers (see ~\cite{Tabor:EPL1986,deGennes:book:Intro,Sreenivasan:JFM2000,Benzi:ARCMP2018} and references therein).  It is a difficult problem due to complex physics of turbulence and polymers.   Despite  many experimental and theoretical attempts, we are far from consensus on the mechanism behind this phenomena.    Researchers attribute the following factors for the drag reduction: viscoelasticity, nonlinear interactions between the polymer and the velocity field, interactions at the boundary layers, anisotropic stress etc.~\cite{Tabor:EPL1986,deGennes:book:Intro,Sreenivasan:JFM2000,Benzi:ARCMP2018}.  Both, bulk and boundary layer dynamics may play a significant role in drag reduction. Yet, researchers believe that the contributions from the bulk probably dominates that from the boundary layer~\cite{Sreenivasan:JFM2000}.  Several experiments and numerical simulations  reveal that bubbles and surfactants too suppress turbulent drag reduction~\cite{Spandan:JFM2018}.

In this paper we present turbulent drag reduction from the perspectives of energy transfers and energy flux in the bulk flow.  In a turbulent flow forced at large scales, the injected  kinetic energy  cascades to intermediate scale, and then to small scales, where the energy flux is dissipated by viscous force.   In pure hydrodynamics turbulence, the energy injection rate, kinetic energy flux, and the viscous dissipation are equal, and it is denoted by   $\Pi_u$~\cite{Kolmogorov:DANS1941Dissipation,Kolmogorov:DANS1941Structure,Davidson:book:Turbulence,Sagaut:book}. The turbulent drag is proportional to the kinetic energy flux.  

In a turbulent flow, in the presence of magnetic field, polymers, or bubbles, a part of  kinetic energy flux is converted to this elastic energy.  In such flows, coiled polymers act like springs; magnetic field act as taut strings; and bubbles act as elastic spheres; hence, they posses elastic energies. The above energy transfers lead to a reduction in kinetic energy flux, and hence in turbulent drag.  In this paper we show that the above generic process is one of the   prime causes of turbulent drag reduction  in magnetohydrodynamics (MHD), quasi-static MHD (QS MHD),  polymeric flows, and bubbly turbulence.

Among a large body of work on turbulent drag reduction in polymers,  works related to energy flux are quite small in number.   Recently, Valente et al.~\citep{Valente:JFM2014,Valente:PF2016} performed numerical simulations of polymeric solution and computed various energy fluxes.  They showed a transfer of kinetic energy to the elastic energy for a set of parameters.    Using numerical simulations, Benzi et al.~\citep{Benzi:PRE2003} and Perlekar et al.~\citep{Perlekar:PRL2006} also analysed energy spectra and dissipation rates of kinetic and elastic energies.  In this paper we invoke some of  these numerical results for our arguments on drag reduction in polymeric turbulence. 

There are a large body of works on energy transfer computations in magetohydrodynamic (MHD) and quasi magnetoydsodynamic (QS MHD) turbulence~\cite{Dar:PD2001,Alexakis:PRE2005,Mininni:PRE2005S2S,Debliquy:PP2005,Kumar:EPL2014,Verma:ROPP2017}.  In these works, for most parameters, there is a preferential energy transfer from kinetic energy to magnetic energy.  These transfers too suppress the kinetic energy flux, that in turn decrease the nonlinearity or turbulent drag compared to hydrodynamic turbulence.  In this paper we present the above results in a common framework of energy flux.   


We consider a general framework for a turbulent flow with a  field ${\bf w}$ embedded in it.  At present, for convenience, we assume ${\bf w}$ to be a vector, but it could  also be a scalar or a tensor.  The equations for the flow are given below~\cite{Fouxon:PF2003,Davidson:book:Turbulence,Sagaut:book}:
\bea
\frac{\partial{\bf u}}{\partial t} + ({\bf u}\cdot\nabla){\bf u}
& = & -\nabla({p}/{\rho}) +  \nu\nabla^2 {\bf u} + {\bf F}_u({\bf u,w}) +  {\bf F}_\mathrm{ext},  \label{eq:U} \nonumber \\ \\
\frac{\partial{\bf w}}{\partial t} + ({\bf u}\cdot\nabla){{\bf w}}
& = &   \eta \nabla^2 {{\bf w}} + {\bf F}_w({\bf u,w}),   \label{eq:W} \\
\nabla \cdot {\bf u}  & = & 0, \label{eq:incompress}
 \eea
 where ${\bf u}, p$ are respectively the velocity and pressure fields; $\rho$ is the density which is assumed to be unity;   $\nu$ is  the kinematic viscosity;  $ \eta$ is the diffusion coefficient for the vector field; and ${\bf F}_u, {\bf F}_w$ are respectively the force fields for the velocity and vector fields arising due to interactions among themselves.   $ {\bf F}_\mathrm{ext}$ is the external field that is employed at large scales to maintain a steady state.
 
\begin{figure}
\centering
\includegraphics[width=1.0\linewidth]{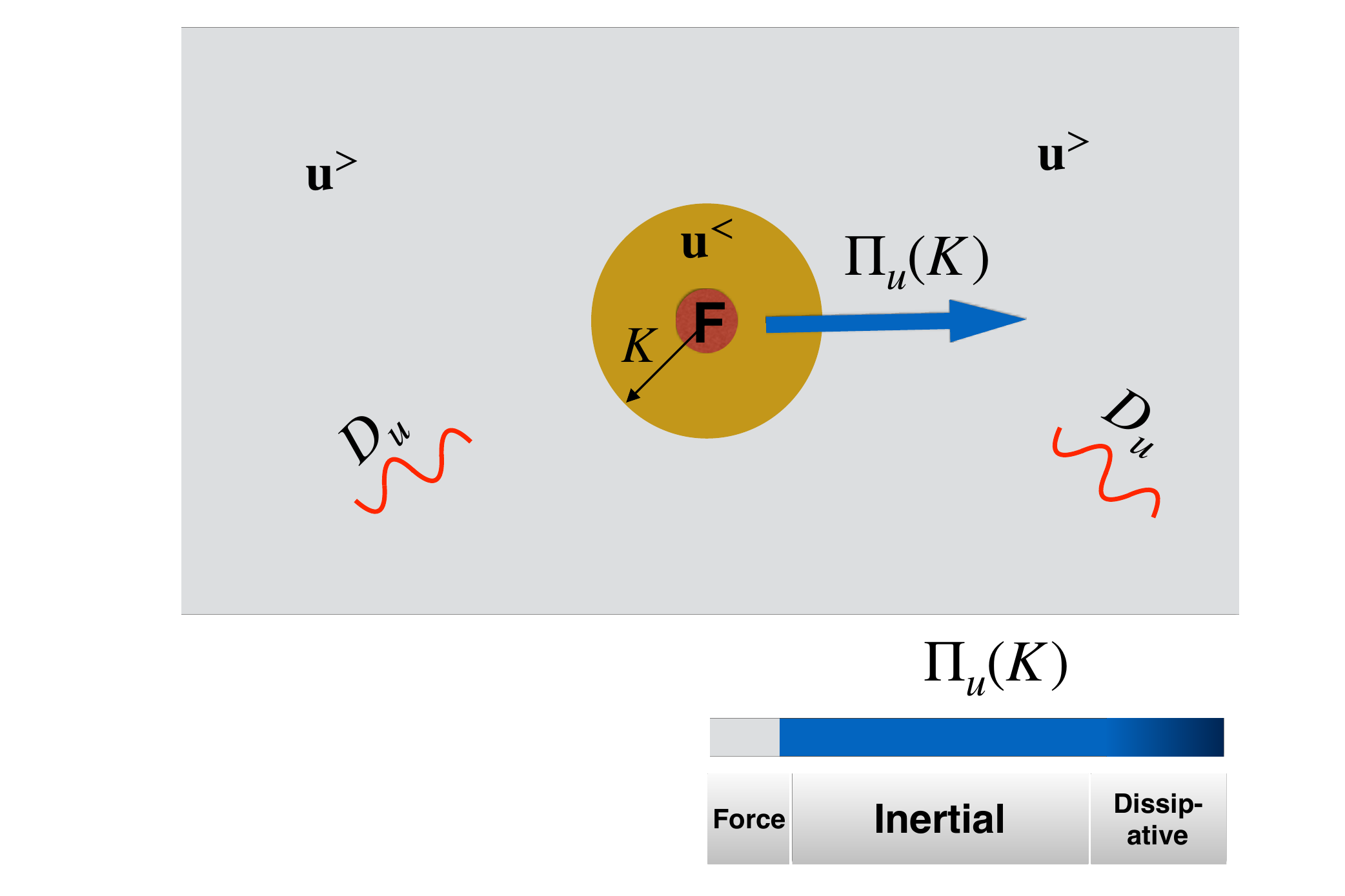}
\caption{(color online) An illustration of kinetic energy flux $\Pi_u(K)$, which is the net  kinetic energy transferred from the modes inside the wavenumber sphere of radius $K$ (yellow sphere) to the modes outside the sphere.   Kinetic energy is injected into the small red sphere of radius $k_f$.  $\Pi_u(K)$, which is constant in the inertial range, is destroyed in the dissipation range via viscous dissipation rate $D_u$.   }
\label{fig:flux}
\end{figure}

 Before venturing into a discussion on mixed turbulence with ${\bf u}$ and ${\bf w}$, we describe  energy flux in Kolmogorov's theory of hydrodynamic turbulence (with  ${\bf w}=0$).   In the inertial range of hydrodynamic turbulence, $\epsilon_\mathrm{inj}$, the energy injected by the external force $\mathcal{F}_\mathrm{ext}$,  cascades to the inertial range as energy flux $ \Pi_u(k) $~\cite{Tabor:EPL1986,deGennes:book:Intro,Sreenivasan:JFM2000,Benzi:ARCMP2018}:
 \be
 \Pi_u(k) \approx     \int_0^{k_f}  d{\bf k} \mathcal{F}_\mathrm{ext}({\bf k}) \approx \epsilon_\mathrm{inj},
 \ee
 where $ \mathcal{F}_\mathrm{ext}({\bf k})  =   \Re[ {\bf F}_\mathrm{ext}({\bf k}) \cdot {\bf u}^*({\bf k})  ]  $.  This energy flux is dissipated in the dissipative range via modal dissipation rate $D_u({\bf k}) = 2 \nu k^2 E_u({\bf k})$.  Hence,
  \be
 \Pi_u(k) \approx     \epsilon_\mathrm{inj} \approx \epsilon_u \approx  \frac{U^3}{d} ,
 \ee
 where $\epsilon_u$ is the total viscous dissipation rate.  We illustrate the kinetic energy flux and the viscous dissipation  in Fig.~\ref{fig:flux}.  
 
 We estimate the energy injection rate by the external force as $F_D U$, where $F_D$ is an estimate of turbulent drag.  Hence, the above equation yields $F_D \approx U^2/d$.  Note that the turbulent drag is determined essentially  by the nonlinear term $({\bf u}\cdot\nabla){\bf u}$, or by the kinetic energy flux as $\Pi_u/U$. 

 Now for the mixed flow with ${\bf u}$ and ${\bf w}$, both the forces, ${\bf F}_u$ and  ${\bf F}_w$, are typically nonlinear.  For example, in MHD where ${\bf w}$ is the magnetic field, ${\bf F}_u = ({\bf \nabla \times w)\times w}$ is the Lorentz force, while  ${\bf F}_w = {\bf (w \cdot \nabla) w}$ represents the stretching of the magnetic field by the velocity field.  Such interactions lead to energy exchanges among the field variables.  It is convenient to represent these transfers in terms of  energy fluxes.  These fluxes are quite complex, and not all of them are relevant here.  In this section we describe the fluxes associated with the velocity field because they are important for the drag reduction.
 
 \begin{figure}
\centering
\includegraphics[width=1.0\linewidth]{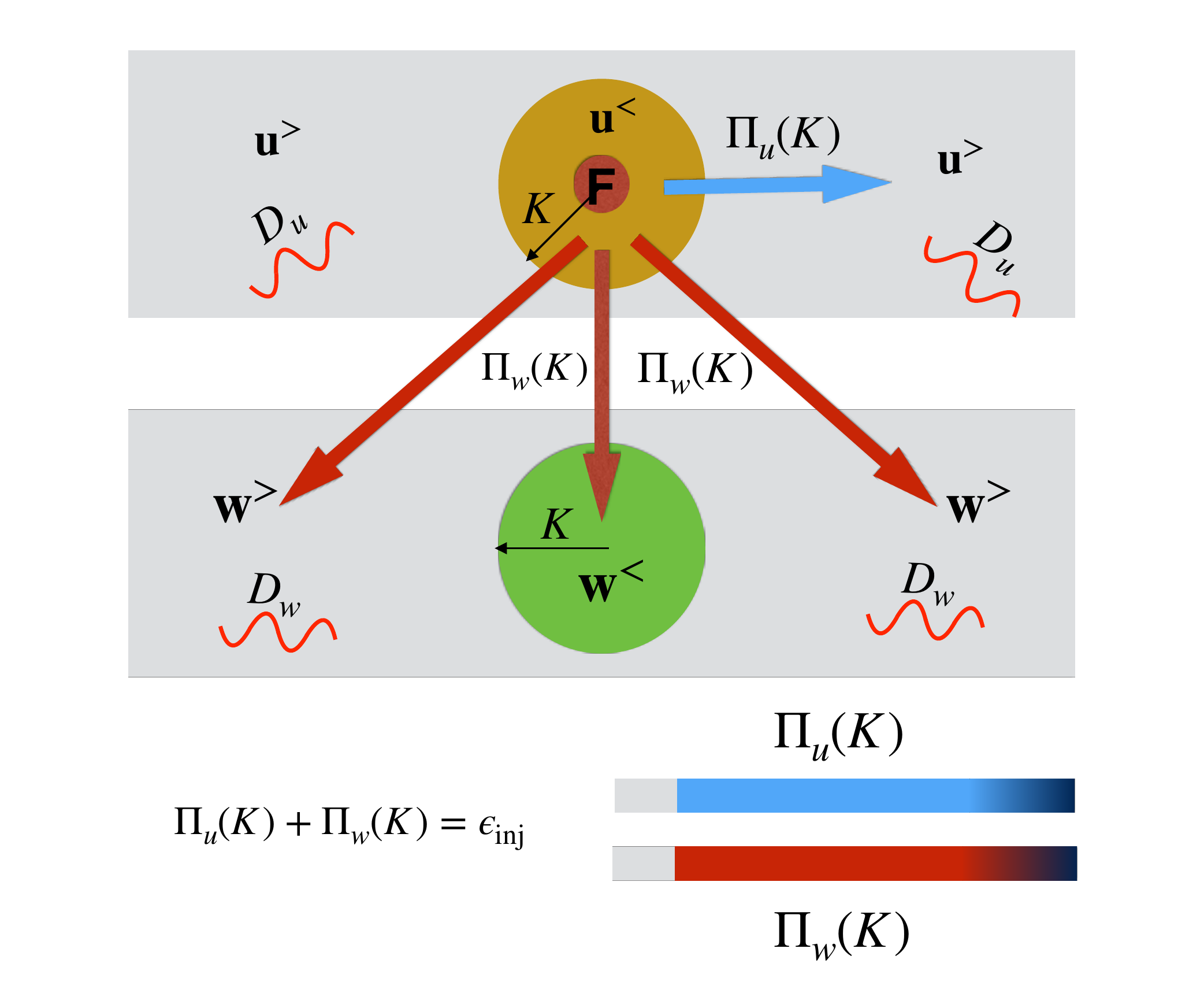}
\caption{(color online) A fraction of  kinetic energy flux is transferred to the field ${\bf w}$.  $\Pi_u(K)$ is the kinetic energy flux for the velocity wavenumber sphere of radius $K$ (yellow sphere), and $\Pi_w(K)$ is the net energy transfer from ${\bf u}$ modes inside the sphere to all the ${\bf w}$ modes.   The   external force injects kinetic energy  into the small red sphere with the rate of $\epsilon_\mathrm{int}$.  The energy fluxes are dissipated by  the dissipation rates $D_u$ and $D_w$.  In the inertial range, $\Pi_u(K) + \Pi_w(K) = \epsilon_\mathrm{int}$.}
\label{fig:flux_w}
\end{figure}
 Computation of multiscale energy transfers are quite convenient in spectral space.  The modal kinetic energy of wavenumber ${\bf k}$ is $E_u({\bf k}) = |{\bf u(k)}|^2/2$.  The evolution equation for the net kinetic energy of a wavenumber sphere of radius $K$ is given by the following equation~\cite{Davidson:book:Turbulence,Sagaut:book,Verma:book:BDF}:
 \bea
 \frac{d}{dt} \int_0^{K} d{\bf k} E_u({\bf k}) &  =  &-  \Pi_u(K) -   \Pi_w(K) +  \int_0^{k_f}  d{\bf k} \mathcal{F}_\mathrm{ext}({\bf k})]    \nonumber \\
 && -  \int_0^{K}  d{\bf k}  D_u({\bf k}) 
 \label{eq:var_Pi}
  \eea 
  where $ \Pi_u(K)$ is kinetic energy flux arising due to $ ({\bf u}\cdot\nabla){\bf u}$ term of Eq.~(\ref{eq:U}); $\Pi_w(K)$ is the energy flux from ${\bf u}$ to ${\bf w}$ due to ${\bf F}_u({\bf u,w}) $;   $D_u({\bf k})$ is the viscous dissipation rate; and $\mathcal{F}_\mathrm{ext}({\bf k})$ is the  energy injection rate by  ${\bf F}_\mathrm{ext}$ that is active in the wavenumber band $[0,k_f]$. We illustrate these energy transfers in Fig.~\ref{fig:flux_w}.  More details are given in Supplemental Material~\cite{SM:PRL}.
  
For the following discussion, we assume that the flow is statistically steady, i.e., $d   \int_0^{K} d{\bf k} E_u({\bf k}) /dt =0$.   For this, in the inertial range where $D_u({\bf k}) =0$, Eq.~(\ref{eq:var_Pi}) yields
 \be
 \Pi_u(k) + \Pi_w(k)  \approx \epsilon_\mathrm{inj}.
 \label{eq:Pi_sum}
 \ee
 See Fig. 2 for an illustration.  For MHD, QS MHD, and polymeric turbulence, it has been observed that $\Pi_w(k)>0$, that is, the velocity field injects energy into the magnetic field in MHD and QS MHD turbulence, or to the polymers in the  polymeric turbulence, or to the bubbles in bubbly turbulence (detailed discussions in subsequent sections).  Therefore, using Eq.~(\ref{eq:Pi_sum}) we deduce that for the same injection rate $\epsilon_\mathrm{inj}$, the kinetic energy flux in the mixture (with field ${\bf w}$) is lower than the corresponding flux in hydrodynamic turbulence.  That is,
 \be
 \Pi_{u,\mathrm{mixture}} <  \Pi_{u,\mathrm{hydro}}.
 \ee 
 Since turbulent drag  $F_D \approx \Pi_u/U$, we expect that
  \be
 F_{D,\mathrm{mixture}} <  F_{D,\mathrm{hydro}}.
 \ee 
Thus, turbulent drag is reduced in the presence of magnetic field, polymers, or bubbles.   Quantitatively, it will be more appropriate to compute a drag-reduction coefficient:
\be
c = \frac{ \Pi_{u,\mathrm{mixture}} }{U^3/d}.
\label{eq:c}
\ee
Note that $c \approx 1$ for hydrodynamic turbulence, and it will be lower than unity in the presence of magnetic field or polymers.

 
For magnetohydrodynamic (MHD) turbulence, the governing equations are Eqs.~(\ref{eq:U}, \ref{eq:W}, \ref{eq:incompress}) with ${\bf w}$  as  the magnetic field; ${\bf F}_u = ({\bf \nabla \times w)\times w}$ is the Lorentz force; and  ${\bf F}_w = {\bf (w \cdot \nabla) u}$ represents stretching of the magnetic field by the flow~\cite{Verma:PR2004}.  The complex set of nonlinearities in the flow yield various energy fluxes~\cite{Verma:PR2004}.  However, for the present discussion, we need to focus only on the kinetic energy flux, $\Pi_u$, and energy flux to the magnetic field, $\Pi_w$.  See Fig.~\ref{fig:flux_w} for an illustration.

 Researchers have studied the energy fluxes $\Pi_u$ and $\Pi_w$ in detail for various combinations of $\nu$ and $\eta$, or their ratio $\mathrm{Pm} = \nu/\eta$, which is called the {\em magnetic Prandtl number}.   Mininni et al.~\citep{Mininni:ApJ2005} computed the fluxes  $\Pi_u$ and $\Pi_w$ using numerical simulations, and observed that   $\Pi_w > 0$, and that the kinetic energy flux $\Pi_u$ of MHD turbulence is smaller than the corresponding $\Pi_u$  of hydrodynamic turbulence, i.e.  
 \be
  \Pi_{u,\mathrm{MHD}} <  \Pi_{u,\mathrm{hydro}}.
  \ee    
  Debliquy et al.~\citep{Debliquy:PP2005}, Kumar et al.~\citep{Kumar:EPL2014},  Verma and Kumar~\citep{Verma:JoT2016}, and other researchers arrived at  similar conclusions.  In particular, Verma and Kumar~\citep{Verma:JoT2016} simulated MHD shell model for Pm = 1 and showed that under a steady state, in the inertial range, $\Pi_u \approx 0.06 \pm 0.02$ and $\Pi_w \approx 0.93 \pm 0.02$; this result indicates a drastic reduction of kinetic energy flux in MHD turbulence.   Note that, $\Pi_w$,  the energy transferred to the magnetic field from the velocity field,  is responsible  for the enhancement of magnetic field in astrophysical dynamos (e.g., planets, stars, and galaxies)~\cite{Moffatt:book}.
  
As argued in the previous section, the depleted $\Pi_u$ in MHD turbulence  leads to a reduction in turbulent drag.  Physically, the magnetic field makes the flow less random (or more ordered) compared to hydrodynamic turbulence.  Therefore, we expect the turbulent drag  in MHD turbulence to be lower than the corresponding hydrodynamic counterpart.  For a more quantitative description, it would be important to compute $c$ of Eq.~(\ref{eq:c}) for MHD turbulence.

Quasi-static  (QS) MHD turbulence, a special class of MHD flows, has very small magnetic Prandtl number~\cite{Knaepen:ARFM2008,Verma:ROPP2017}.    Such flows are observed in liquid metals with a strong external magnetic field. Here, the Lorentz force is proportional to $-N {\bf u }$, hence it is dissipative.  The parameter $N$, called {\em interaction parameter}, is the ratio of Lorents force and nonlinear term ${\bf (u \cdot  \nabla) u}$.   This dissipative force transfers the kinetic energy to the magnetic energy, which is immediately destroyed by Joule dissipation.  QS MHD turbulence  models of Verma and Reddy~\citep{Verma:PF2015QSMHD} clearly show a reduction in the kinetic energy flux with the increase of $N$. This suppression of  kinetic energy flux $\Pi_u$ leads to reductions in  the nonlinearity (${\bf (u \cdot \nabla) u}$) or  in turbulent drag.

In addition, Reddy and Verma~\cite{Reddy:PF2014} simulated QS MHD turbulence for a wide range of interaction parameter $N$.  In Table~\ref{tab:QSMHD} we list  the rms velocity $U$ as a function of $N$  for their runs  with a constant  energy injection rate of 0.1 (in nondimensional unit).   Here, $U$ is measured in units of $L/T$, where $L,T$ are length and time scale of large scale eddies.  Clearly,  $U$ increases monotonically with $N$.  In other words,  the flow becomes more and more ordered with the increase of $N$ that leads to reductions in nonlinearity and turbulent drag.   It is important to note that large $U$ does not imply larger nonlinearity (${\bf (u \cdot \nabla) u}$), which  depends on $U$, as well as on the phase relations between the velocity modes.  The magnetic field alters the phase relations in ${\bf (u \cdot \nabla) u}$; suppresses nonlinearity and drag; and produces larger $U$.

 \begin{table}
\caption{For simulation of QS MHD turbulence by Verma and Reddy~\cite{Verma:PF2015QSMHD}, root mean square (rms) velocity of the flow as a function of interaction parameter $N$.  The flow speed increases with the increase of $N$. }
\begin{tabular}{l  | r}
\hline
$N$ & $U$  \\
\hline
1.7 & 0.39	 \\ \hline
18 & 0.51 \\ \hline
27 & 0.65 \\  \hline
220 & 0.87 \\  \hline
\end{tabular}
\label{tab:QSMHD}
\end{table}

Turbulence plays an important role in drag reduction. In contrast, a laminar QS MHD flow exhibits stronger drag than its hydrodynamic counterpart (with ${\bf B}_0=0$).  That is, the velocity in laminar QS MHD is lower than in laminar hydrodynamics~\citep{Moreau:book:MHD,Verma:ROPP2017}.  Hence, it is the suppression of kinetic energy flux that is responsible for drag reduction with magnetic field.  Also note  that walls play an important role in QS MHD; these effects however are beyond the scope of this paper.

Thus, MHD and QS MHD turbulence illustrate how inclusion of magnetic field in the flow leads to a suppression of  kinetic energy flux, and hence a reduction in  turbulent drag.  This finding may be useful for  engineering applications involving liquid metals.  Next, we show that a similar process is at work in turbulent flows with dilute polymers.
 
 
 The equations for a turbulent flow with dilute polymers is similar to Eqs.~(\ref{eq:U}, \ref{eq:W}, \ref{eq:incompress}), except that ${\bf w}$ is replaced by a tensor field $\mathcal{C}$ representing a polymer.  In particular, we focus on a polymer that is represented by the finitely extensible nonlinear elastic-Peterlin (FENE-P) model~\cite{Sagaut:book,Perlekar:PRL2006}.  
 
 In FENE-P model, ${\bf F}_u =  \nabla \cdot (f \mathcal{C})$ and $F_\mathcal{C} =  (\nabla {\bf u})^T \cdot \mathcal{C} +  \mathcal{C} \cdot  (\nabla {\bf u}) $, where $f$ is a function of  $\mathcal{C}$.  The tensorial nature of $\mathcal{C}$ makes the physics more complex~\cite{Sagaut:book}. Yet, energetics arguments provide a schematic picture of energy fluxes and drag reduction.  These arguments are somewhat independent of detailed dissipation mechanism.  As described in the introduction, drag reduction in polymeric solution depends on boundary layers,  bulk dynamics, anisotropy, etc.  However, as indicated by many researchers~\cite{Tabor:EPL1986,deGennes:book:Intro,Sreenivasan:JFM2000,Benzi:ARCMP2018}, transfers from kinetic energy to elastic energy play a major role in drag reduction.    The term ${\bf F}_u \cdot {\bf u}$ yields $\Pi_\mathcal{C}>0$ (corresponding to $\Pi_w$ of Eq.~(\ref{eq:var_Pi})), or a net transfer of kinetic energy to elastic energy of the polymer.  Supplemental Material~\cite{SM:PRL} contains a detailed description of the above equations and terms.

Fouxon and Lebedev~\citep{Fouxon:PF2003} showed that the equations for dilute polymers are intimately connected to those of MHD turbulence.  Hence, we expect that the energy transfers in polymeric turbulence to be similar to those of MHD turbulence.   Using numerical simulations, Valente et al.~\citep{Valente:JFM2014,Valente:PF2016}  analysed the energy transfers, in particular fluxes $\Pi_u$ and $\Pi_\mathcal{C}$,  for polymeric turbulence.  The energy fluxes $\Pi_u$, $\Pi_\mathcal{C}$ depend on the Deborah number, $\mathrm{De}$,  which is the ratio of the relaxation  time scale of the polymer and the characteristic time scale for the energy cascade.  A common feature among all the numerical runs  is that  $\Pi_\mathcal{C} >0$, and that $\Pi_u$ is always reduced, but the energy transfer from kinetic to elastic is maximum when $\mathrm{De} \sim 1$. For example, Valente et al.~\cite{Valente:JFM2014} showed that for $\mathrm{De} = 1.17$, $\Pi_\mathcal{C}/\epsilon_\mathrm{inj} \approx 0.8$ for $k\eta > 0.2$ where $\eta$ is Kolmogorov's wavenumber.  On the other hand $\Pi_u/\epsilon_\mathrm{inj} \approx 0.1$ near $k\eta \approx 0.2$ and $\approx 0$ for other wavenumbers.  The balance, $0.2 \epsilon_\mathrm{inj}$ is dissipated by viscosity.  Thus, $\Pi_u$ is drastically reduced in the presence of polymers. 

Thus, analogous to MHD turbulence, flows with dilute polymer also  exhibit reduction in kinetic energy flux due to the transfer of kinetic energy to elastic energy.  That is,
 \be
  \Pi_{u,\mathrm{Polymeric}} <  \Pi_{u,\mathrm{hydro}}.
  \ee   
 This reduction leads to a decrease in nonlinearity, and hence turbulent drag.   In bubbly turbulence, we expect turbulence to facilitate  transfers from kinetic energy to elastic energy of the bubbles that may be treated like elastic spheres.   Researchers~(e.g.~\cite{Spandan:JFM2018})  argue that bubbles induce drag reduction in turbulence due to the above broader analogy between polymers  and bubbles.
 

Now we summarise tour results.  Turbulent drag reduction is an important problem of science and engineering.  In this paper, using a general framework, we show that  drag reduction is due to a partial transfer of kinetic energy flux to the secondary field, such as magnetic field, polymer, or bubbles.  This transfer leads to a suppression of nonlinearity.    We quantify our claim using the past results on energy fluxes in magnetohydrodynamics, quasi-static magnetohydrodynamics, and polymeric turbulence.  Our results are consistent with earlier works on drag reduction in  polymer solution.  This picture predicts drag reduction in MHD and QS MHD turbulence.

The formulation presented in the paper provides valuable insights in the mechanism of turbulent drag reduction, as well as  measures for the quantification of the reduced drag.  These results also indicate that controlled magnetic field can be used for drag reduction in liquid metal flows.


\acknowledgements
The author thanks Abhishek Kumar, Franck Plunian, Shashwat Bhattacharya, and Supratik Banerjee for useful  discussions. This work was supported by the Indo-Russian project (DST-RSF) INT/RUS/ RSF/P-03 and RSF-16-41-02012;


\pagebreak

{\bf SUPPLEMENTAL MATERIAL}
\vspace{0.3cm}
\section*{Equations for the energy and fluxes}
The flow equations with a vector field ${\bf w}$ are~\cite{Fouxon:PF2003,Davidson:book:Turbulence,Sagaut:book}
\bea
\frac{\partial{\bf u}}{\partial t} + ({\bf u}\cdot\nabla){\bf u}
& = & -\nabla({p}/{\rho}) +  \nu\nabla^2 {\bf u} + {\bf F}_u({\bf u,w}) +  {\bf F}_\mathrm{ext},  \label{eq:U} \nonumber \\ \\
\frac{\partial{\bf w}}{\partial t} + ({\bf u}\cdot\nabla){{\bf w}}
& = &   \eta \nabla^2 {{\bf w}} + {\bf F}_w({\bf u,w}),   \label{eq:W} \\
\nabla \cdot {\bf u}  & = & 0, \label{eq:incompress}
 \eea
 where ${\bf u}, p, \rho$ are respectively the velocity,  pressure, and density fields; $\nu$ is  the kinematic viscosity;  $ \eta$ is the diffusion coefficient for the vector field;  ${\bf F}_u, {\bf F}_w$ are respectively the force fields for the velocity and vector fields due to internal interactions; and $ {\bf F}_\mathrm{ext}$ is the external field at the large scales.  Using Eq.~(\ref{eq:U})  we derive the following equation for the kinetic energy density $u^2/2$:
\be
\frac{\partial}{\partial t} \frac{u^2}{2} +  \nabla \cdot \left[ \frac{u^2}{2} {\bf u} \right] = - \nabla \cdot ( p {\bf u}) + [{\bf F}_u + {\bf F}_\mathrm{ext} ] \cdot {\bf u} -\nu {\bf u} \cdot  \nabla^2 {\bf u}.
\label{eq:basic_hd:framework:Eu_dynamics2}
\ee

In Fourier space, the corresponding equation for the modal kinetic energy  $E_u({\bf k}) = |{\bf u(k)}|^2/2$ is 
 \bea
  \frac{d}{dt} E_u(\mathbf{k})  & = & T_{u}({\bf k})  +  \mathcal{F}_u({\bf k})  + \mathcal{F}_\mathrm{ext}({\bf k})-D_u(\mathbf{k}),
 \label{eq:MHD_ET:Eu_dot_Fext} 
\eea
where 
 \bea
 T_u({\bf k}) & = & \sum_{\bf p} \Im \left[ {\bf  \{  k \cdot u(q) \} \{ u(p) \cdot u^*(k) \} } \right] , \\
 \mathcal{F}_u({\bf k}) & =  & \Re[ {\bf F}_u({\bf k}) \cdot {\bf u}^*({\bf k})  ], \\
 \mathcal{F}_\mathrm{ext}({\bf k}) & = &   \Re[ {\bf F}_\mathrm{ext}({\bf k}) \cdot {\bf u}^*({\bf k})  ], \\
 D_u(\mathbf{k}) & = & -2 \nu k^2 E_u({\bf k}),
 \eea
 with $\Re, \Im$ representing the real and imaginary parts respectively, and ${\bf q=k-p}$.  In this paper we do not discuss the energetics of ${\bf w}$ field because the turbulent drag reduction is related to the energy fluxes associated with the velocity field.  When we sum the above equation over all the modes in the wavenumber sphere of radius $K$, we obtain the following equation~\cite{Davidson:book:Turbulence,Sagaut:book,Verma:book:BDF}:
 \bea
 \frac{d}{dt} \sum_{k \le K}  E_u({\bf k})   &= &  \sum_{k \le K} T_u({\bf k}) + \sum_{k \le K}\mathcal{F}_u({\bf k})  \nonumber \\
 && + \sum_{k \le K}\mathcal{F}_\mathrm{ext}({\bf k}) - \sum_{k \le K} D_u({\bf k}). 
 \label{eq:ET:Pi_k0_from_Ek}
  \eea  
  Physical interpretations of the four terms in the right-hand side of Eq.~(\ref{eq:ET:Pi_k0_from_Ek}) are as follows:
  \begin{enumerate}
  \item $\sum_{k\le K} T_u({\bf k})$ is the net energy transfer from the modes outside the sphere to the modes inside the sphere due to the nonlinearity $ ({\bf u}\cdot\nabla){\bf u}$.
  \item $\sum_{k\le K} \mathcal{F}_u({\bf k}) $ is the total energy transfer rate inside the sphere by the interaction force ${\bf F}_u({\bf k})$.
  \item $\sum_{k\le K} \mathcal{F}_\mathrm{ext}({\bf k}) $ is the net energy injected by the external force ${\bf F}_\mathrm{ext}$ (red sphere of Fig.~2 of main text).  For any $K$ beyond the forcing band, the above sum is the net energy injection rate $\epsilon_\mathrm{inj}$ by the external force.
\item $\sum_{k\le K} D_u({\bf k})$ is the total viscous dissipation rate inside the sphere.
\end{enumerate}
  
The kinetic energy flux $\Pi_u(K)$ is defined as the cumulative kinetic energy transfer rate from ${\bf u}$ modes inside the sphere to ${\bf u}$  modes outside the sphere.  See Figures.~1 and 2 of the main text for an illustration.  In terms of Fourier modes, we compute this flux as
\bea
\Pi_u(K) &= & \sum_{p\le K} \sum_{k>K} \Im \left[ {\bf  \{  k \cdot u(q) \} \{ u(p) \cdot u^*(k) \} } \right] 
 \nonumber \\
 &= &  - \sum_{k\le K} T_u({\bf k}) .
\label{eq:ET:fluid_flux}
\eea
 Note that $-\sum_{k\le K} \mathcal{F}_u({\bf k}) $ represents the net energy transfer from the ${\bf u}$ modes inside the sphere to all the ${\bf w}$ modes due to the interacting force $\mathbf{F}_u({\bf k})$. Hence we define the corresponding flux $\Pi_w(K)$ as
 \be
 \Pi_w(K)=-\sum_{k\le K} \mathcal{F}_u({\bf k}) .
 \ee
 Under a steady state, for any  wavenumber sphere, the kinetic energy injected  by ${\bf F}_\mathrm{ext}$ is lost due to the two flux $\Pi_u$, $ \Pi_w$, and the viscous dissipation rate. That is, 
 \be
 \Pi_u(K) + \Pi_w(K) + \sum_{k \le K} D_u({\bf k}) = \epsilon_\mathrm{inj}.
 \ee
 In the inertial range, $D_u({\bf k}) \approx 0$ that leads to
  \be
 \Pi_u(K) + \Pi_w(K)  \approx \epsilon_\mathrm{inj}.
 \label{eq:Pi_sum}
 \ee
 
\section*{Energetics of MHD and QS MHD turbulence}
 
 For MHD turbulence, ${\bf w}$ is the magnetic field, and  $ {\bf F}_u = {\bf (w \cdot \nabla) w}$ is the Lorentz force.  In Fourier space, $ {\bf F}_u$   and the corresponding energy transfer rate are~\cite{Verma:PR2004,Dar:PD2001}
 \bea
 {\bf F}_u({\bf k})  & =  & i  {\bf  \{  k \cdot w(q) \} w(p)}, \\
  \mathcal{F}_u({\bf k}) & = &  \Re[ {\bf F}_u({\bf k}) \cdot {\bf u^*(k)}] \nonumber \\
  &= &  \sum_{\bf p} - \Im \left[ {\bf  \{  k \cdot w(q) \} \{ w(p) \cdot u^*(k) \} } \right],
  \label{eq:Sbb_sum}
 \eea
 where ${\bf q=k-p}$. 
 The term inside the sum of Eq.~(\ref{eq:Sbb_sum})  is the mode-to-mode energy transfer from ${\bf w(p)}$ to ${\bf u(k)}$ with the mediation of ${\bf w(q)}$, and it is denoted by $S^{ww}({\bf k|p|q})$ \citep{Dar:PD2001,Verma:PR2004}.  Therefore, the corresponding energy flux is
 \bea
 \Pi_w(K) &= & \sum_{k \le K}  \sum_{\bf p} S^{ww}({\bf k|p|q})  \nonumber \\
 & = & \sum_{k \le K}  \sum_{p \le K}  S^{ww}({\bf k|p|q})  + \sum_{k \le K}  \sum_{p > K}  S^{ww}({\bf k|p|q}) \nonumber \\
 & =& \Pi^{u<}_{w<} + \Pi^{u<}_{w>} .
 \eea
 In the above equation, $\Pi^{u<}_{w<} $ represents the energy transfer from the velocity modes inside the sphere (${\bf u}^<$) to the magnetic modes inside the sphere (${\bf w}^<$), while  $ \Pi^{u<}_{w>} $ represents energy transfers from ${\bf u}^<$ modes to ${\bf w}^>$ modes.  The former flux is illustrated by the central red arrow of Fig.~2, while the latter flux by the other two arrows.  Numerical simulations and experiments reveal that $ \Pi_w(K) >0$.  Therefore, using Eq.~(\ref{eq:Pi_sum}) we deduce that
 \be
  \Pi_{u,\mathrm{MHD}} <  \Pi_{u,\mathrm{hydro}}
  \ee   
that leads to a drag reduction. 

In QS MHD turbulence~\cite{Knaepen:ARFM2008,Verma:ROPP2017},  the corresponding force and related energy transfer rates are
 \bea
 {\bf F}_u({\bf k})  & =  &  -N \cos^2 \theta {\bf u(k)} \\
  \mathcal{F}_u({\bf k}) & = &  \Re[ {\bf F}_u({\bf k}) \cdot {\bf u^*(k)}] = - 2 N \cos^2 \theta E_u({\bf k})  ,
 \eea
where $N$ is the interaction parameter, and $\theta$ is the angle between the external magnetic field and the wavenumber ${\bf k}$.  Therefore,  $ \Pi_w(K)$ takes the following form
 \be
 \Pi_w(K)=-\sum_{k\le K} \mathcal{F}_u({\bf k}) = \sum_{k\le K}  2 N \cos^2 \theta E_u({\bf k}) .
 \ee
 Hence, as argued above, the kinetic energy flux is suppressed compared to the hydrodynamic turbulence.

  \section*{Energetics of  turbulence in polymer solution}
 For the polymer solution, under FENE-P approximation, the force $ {\bf F}_u $ and the corresponding energy injection rate are~\cite{deGennes:book:Intro,Perlekar:PRL2006,Benzi:ARCMP2018}
 \bea
  F_{u,i} & = & [\nabla \cdot \mathcal{C}]_i =  \partial_j (f \mathcal{C}_{ij}), \\
   F_{u,i}({\bf k}) & = &  \sum_{\bf p} \Im \left[  k_j f({\bf q}) \mathcal{C}_{ij}({\bf p} \right], \\
   \mathcal{F}_u({\bf k}) & = & \Re[F_{u,i}({\bf k}) u_i^*({\bf k})] \nonumber \\
   & = &   -c_1 \sum_{\bf p} \Im \left[  k_j f({\bf q}) \mathcal{C}_{ij}({\bf p})u^*_i({\bf k}) \right],
  \eea
 where $\mathcal{C}$ is the correlation tensor, and ${\bf q=k-p}$.  The field   $\mathcal{C}$  replaces  ${\bf w}$ of Eqs. (1,2). Given the above, we define the flux corresponding to $\Pi_w(K)$ of Fig. 2 of the main text as
 \bea
 \Pi_\mathcal{C}(K) &= & \sum_{k \le K}  \sum_{p \le K}  -c_1 \Im \left[  k_j f({\bf q}) \mathcal{C}_{ij}({\bf p})u^*_i({\bf k}) \right]\nonumber \\
 & &  + \sum_{k \le K}  \sum_{p > K}  -c_1 \Im \left[  k_j f({\bf q}) \mathcal{C}_{ij}({\bf p})u^*_i({\bf k})  \right]  \nonumber \\
 & = & \Pi^{u<}_{\mathrm{C}<} + \Pi^{u<}_{\mathrm{C}>} .
 \eea
 In the above equation, $\Pi^{u<}_{\mathrm{C}<} $ represents the kinetic to elastic energy transfer within the sphere, while $\Pi^{u<}_{\mathrm{C}>} $ represents the energy transfer from the velocity modes inside sphere to the polymer modes outside the sphere (see Fig.~2 of the main text).  As described in the main text, numerical simulations reveal that $ \Pi_\mathrm{C}(K) >0$, that is, kinetic energy is transferred to the elastic energy.   Therefore, Eq.~(\ref{eq:Pi_sum}) yields
 \be
  \Pi_{u,\mathrm{Polymeric}} <  \Pi_{u,\mathrm{hydro}}
  \ee   
that leads to a drag reduction in the flow.

\end{document}